# Transient Gas-Dynamics Filamentation of High-Power Femtosecond Laser Pulse in Compressed Argon

Yu.E. Geints, P.V. Babushkin, A.M. Kabanov, V.K. Oshlakov, E.E. Khoroshaeva

*V. E. Zuev Institute of Atmospheric Optics SB RAS, 1, Acad. Zuev Square 1, Tomsk 634055, Russia*
*Corresponding author e-mail: ygeints@iao.ru

## Abstract

Filamentation is the primary mode of propagation for high-power femtosecond laser pulses in gases, where the optical nonlinearity of the medium manifests itself on the largest scale. Optical filaments are characterized by high intensity and an anomalously broad spectrum (supercontinuum), the control of which presents a significant challenge. The turbulence of the gaseous medium drastically alters the conditions of laser pulse filamentation and the parameters of the generated supercontinuum due to enhanced stochastic fragmentation of laser beam and earlier arrest of self-focusing. We have experimentally investigated the spectral characteristics and spatial structure of femtosecond pulses from a titanium-sapphire laser during filamentation in an optical cell filled with argon at pressures up to 40 atm under pressure shock-drop conditions. This leads to the development of strong jet flows and vortex gas turbulence, which in turn triggers the early onset of multiple filamentation of the optical pulse and large-scale broadening of its spectrum throughout the entire duration of the pressure drop. The magnitude of this spectrum broadening can reach 80 nm and is proportional to the initial gas pressure. Using computational fluid dynamics simulations, we studied the dynamics of the emergence, development, and relaxation of stimulated turbulence in compressed gas in the region of the cell's outlet valve and assessed the effect it exerts on the propagating femtosecond pulse. The revealed regularities may serve as the basis for developing an effective method of controlling the spectrum of supercontinuum radiation via filamentation of high-power ultrashort laser pulses in gas cells under shock pressure release and rise conditions.

## 1. Introduction

Filamentation of high-power femtosecond laser pulses in gases is the primary propagation mode and represents a complex nonlinear optical process [1-4], which is significantly influenced by random medium inhomogeneities, particularly due to its turbulence. Turbulence introduces stochastic fluctuations in the refractive index of gaseous medium, which drastically changes the conditions for high-intensity radiation propagation and filament formation [5]. Filamentation in a turbulent atmosphere has been actively studied since the 2000s, and significant progress has been made in understanding the mechanisms of turbulence influence and developing methods for controlling the optical filamentation. For instance, as reported in several works [6-12], the distance at which filamentation begins, the number of filaments formed, and their mutual positioning (pointing) become random following Rayleigh statistics. Moreover, depending on the pulse energy and the strength of atmospheric turbulence, the filamentation distance can either increase (global beam self-focusing [13]) or decrease (small-scale self-focusing [8]). However, the already formed filaments turn out to be resistant to stochastic distortions of the laser pulse when passing through a perturbed region with turbulence levels even significantly higher than those observed in the atmosphere [14, 15].

During the filamentation, femtosecond radiation significantly enriches its frequency spectrum and becomes a source of broadband coherent radiation — the supercontinuum. This offers advantages for use in a number of atmospheric-optical applications, particularly for remote detection and physicochemical analysis of atmospheric aerosols of various natures [16]. Supercontinuum generation in gases is of considerable practical interest because it allows spectrum control not only by varying the parameters of the laser pulse, but also by adjusting the density of the gaseous medium. The efficiency of supercontinuum generation in high-pressure gases exhibits a significant dependence on both the pressure (density) and the type of gas used in the optical cell. The spectrum of the broadband radiation typically broadens toward the blue wing due to the increased rate of plasma formation in dense gas [17-21]. Moreover, at sufficiently high pressures (> 30 atm), the spectral width of the supercontinuum reaches saturation [20]. For certain molecular gases ($N_2$, $CO_2$), a tendency toward spectral narrowing is even observed [21].

Principal challenge in applying the supercontinuum generation technique using a high-pressure gas cell is the necessity of using a focused laser pulse to ensure reliable filamentation precisely inside the cell. This involves the use either employing narrow optical beams with low power to reduce the self-focusing distance or utilizing costly focusing optics with high beam resistance when working with wide-aperture radiation featuring high pulse energy. It is in the latter case that it becomes possible to obtain a powerful, highly directional coherent supercontinuum radiation at the cell's output.

In this work, we propose a method to overcome this problem by artificially initiating shock perturbations in the working gas. This leads to intense vortex turbulence inside the cell and triggers small-scale self-focusing of the optical radiation, effectively shifting the onset of filamentation into the volume of compressed gas. Essentially, there is a certain analogy here with the fluid dynamics of shock tubes, in which surfaces with discontinuities in the fluid dynamics parameters of the working medium are created and then shock-destroyed. This process generates strong transient perturbations, accompanied by the formation of compression shock waves and rarefaction regions [22]. Below, we present the results of our experimental demonstration of the proposed method in a 2 m gas cell filled with compressed argon, supported by fluid dynamics modelling.

We show that during a shock pressure drop, a sufficiently wide (2.5 cm) and high-power (200 GW) femtosecond radiation from a Ti:Sapphire laser can nonlinearly transform into a broad supercontinuum pulse with a spectral width of approximately 80 nm.

## 2. Experimental Setup

The experimental setup (Fig. 1) is designed to study supercontinuum generation via femtosecond laser filamentation in high-pressure gas. The laser source is a terawatt femtosecond laser system (Avesta Project Ltd.) based on a titanium-sapphire crystal. The system delivers pulses with the following parameters: repetition rate: 10 Hz, central wavelength $\lambda_0 = 800$ nm, pulse duration (FWHM) 55 fs, pulse energy 10 mJ, beam diameter (at $1/e^2$ intensity level) 25 mm, peak pulse power $P_0 = 170$ GW. The laser beam is directed into a high-pressure gas cell filled with argon via a pivoting mirror.

The optical cell is fabricated from a stainless-steel pipe with an inner diameter of 50 mm, equipped with flanges welded using argon arc welding. The flanges house optical windows: an input crystalline quartz input window (25 mm thick) and an output leucosapphire output window (8 mm thick), both coated with an anti-reflection layer. The cell is mounted on a movable base. The radiation exiting the cell is either projected onto a paper screen with the beam profile recorded using an Andor Clara CCD camera and a Pentax K3 camera, or collected by a lens into an integrating sphere (FOIS-1) for spectral measurements and analyzed using a Maya 2000 Pro fiber spectrometer (Ocean Optics). The gas pressure is monitored using a calibrated A-Flow pressure gauge with an accuracy class of 1.6. The optical cell was filled with high-purity argon gas (volume fraction ≥ 99,995%) at pressures ranging from 1 to 40 atm (tested up to 70 atm).

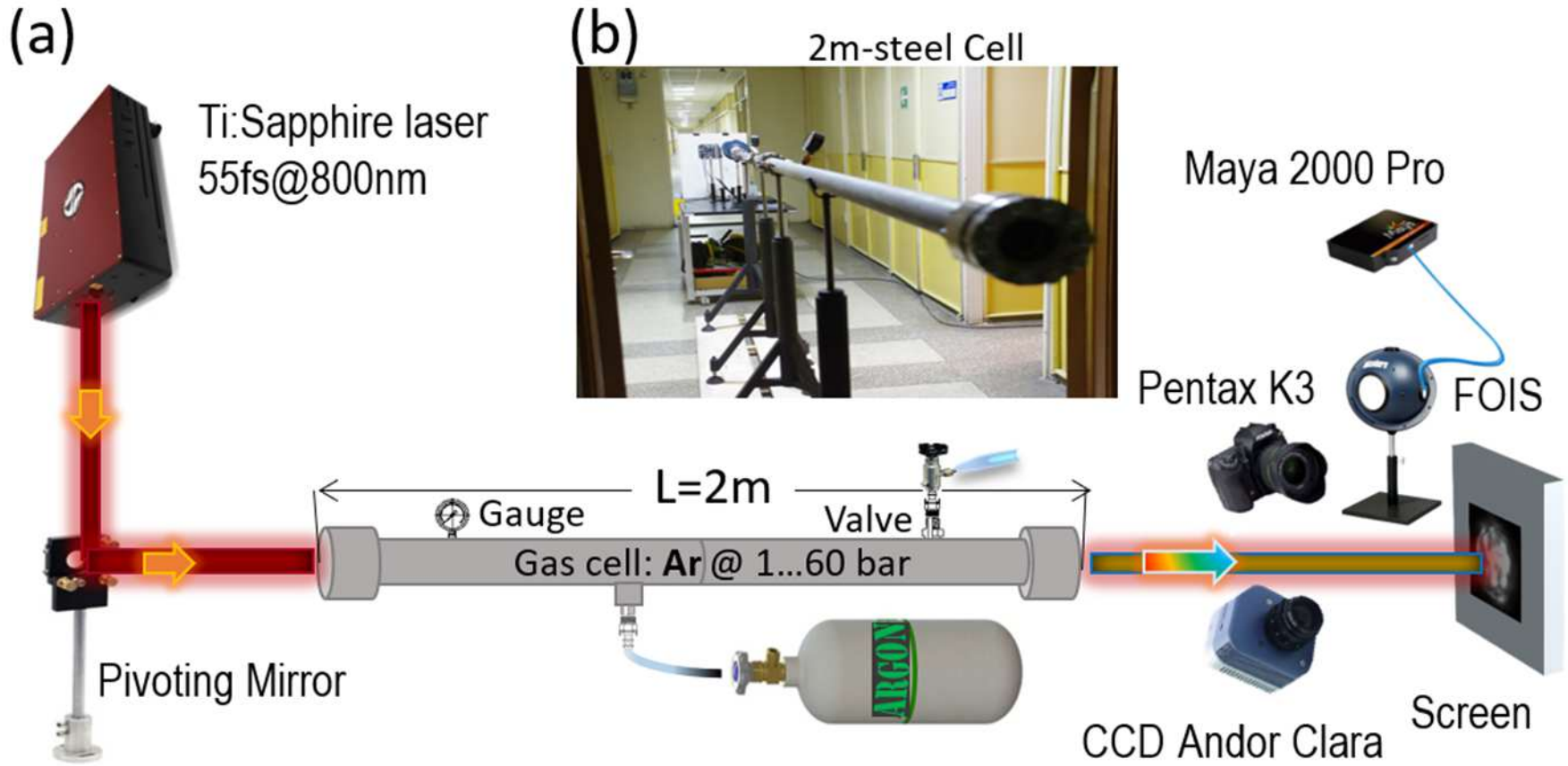

Fig. 1. (a) Experimental setup on shock-dynamic pulse filamentation in compressed gas. (b) An image of a high-pressure steel cuvette with a length of $L = 2$ m.

Spatial and spectral transformations of a femtosecond pulse during filamentation in argon under shock pressure release are shown in Fig. 2(a-d). When processing each spectrum, the data on the spectral composition of the pulse was pre-calibrated according to the transmission function of the spectrometer, and then normalized to the maximum value realized for this particular gas pressure. In addition to the distribution of the spectral power of the radiation, the RMS (effective) pulse spectrum bandwidth (Δλ) is also calculated as the centered second order moment from the square of the spectral power modulus $|U_\lambda|^2$ using the following formula:

$$\Delta\lambda = 2\left[\int |U_\lambda|^2 \left(\lambda^2 - \lambda_g^2\right) d\lambda \Big/ \int |U_\lambda|^2 \, d\lambda\right]^{1/2} \quad (1),$$

where $\lambda_g$ is the spectrum centroid. Note, employing the RMS half-width for characterizing the pulse spectrum proves to be beneficial, as this parameter is independent of the particular form of the spectral function and serves as a universal metric for assessing integral changes in the spectrum.

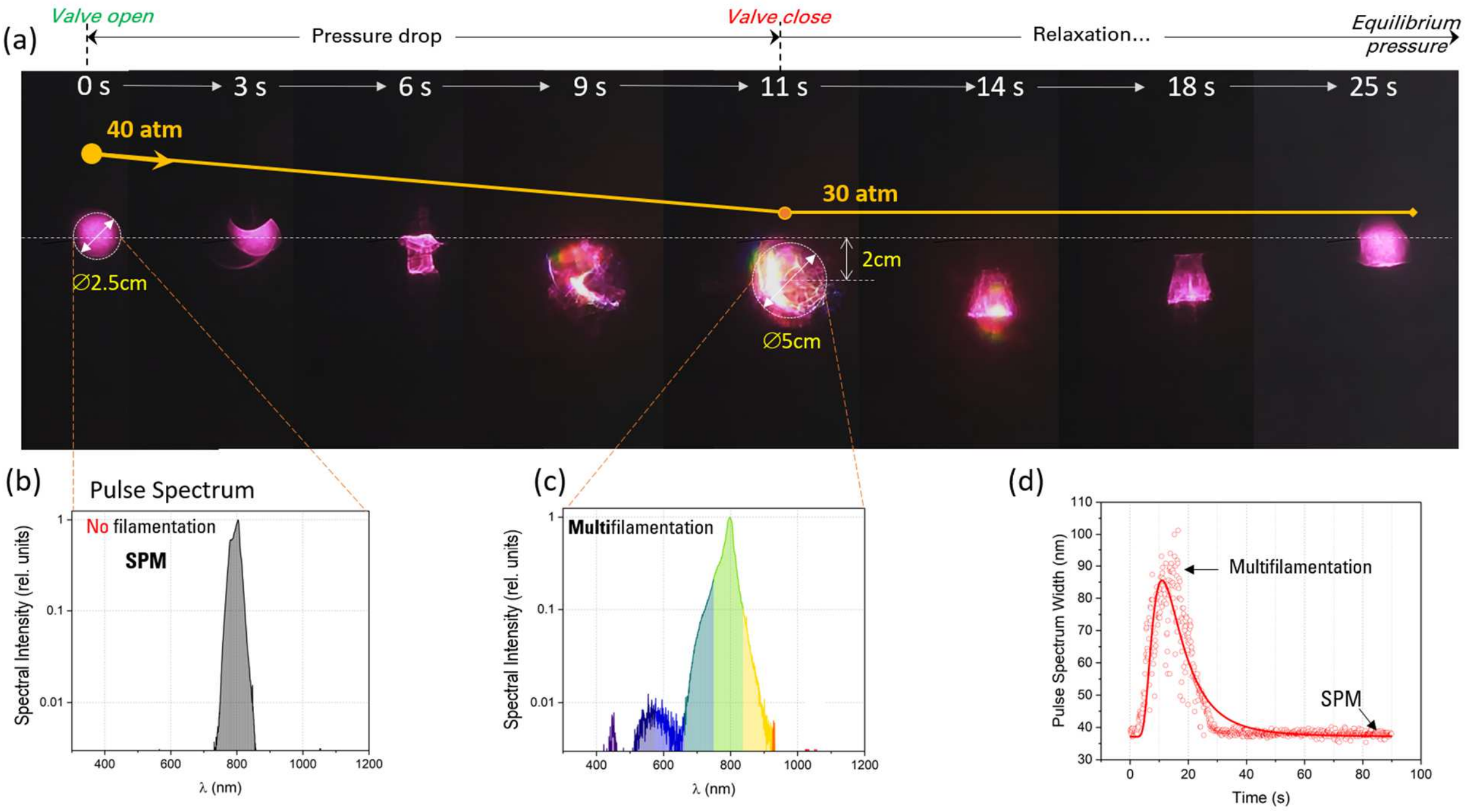

Fig. 2. Characterization of the laser pulse following filamentation in high-pressure argon. (a) Real-time beam profile evolution on the projection screen. Horizontal dashed line marks initial beam position. (b, c) Pulse spectrum at the onset (b) and end (c) of the pressure drop in the cell. (d) Temporal evolution of the effective spectral bandwidth, demonstrating broadband supercontinuum generation. Solid line is the Lorentz function fit.

## 3. Results and discussion

We discuss the temporal dynamics of the transverse profile of the femtosecond beam, presented in Fig. 2a as a sequence of images on a screen placed approximately 0.5 m away from the cell's output window. These images reveal that during the pressure release from the cell, the femtosecond beam exhibits several key changes. First, the beam shifts downward on the screen. Second, it alters its spatial structure, shape, and transverse size. Specifically, it demonstrates the emergence of highly localized inhomogeneities (filament trails), shape aberrations, and an overall increase in beam size. Third, the beam spectral content expands, manifested as a change in color.

The latter effect is clearly observed in panels (b) and (c) of Fig. 2, which show the normalized pulse spectrum at the onset of pressure release and at valve closure (11 s later), respectively. As seen in the data, propagation of a sufficiently high-power laser pulse (≈200 GW) through a 2 m layer of argon compressed to 40 atm does not lead to filamentation of optical radiation. At least, no signature of supercontinuum generation which is typical for filamentation is observed in the pulse spectrum. Instead, only a symmetric broadening of the spectral profile is noted, attributed to self-phase modulation (SPM) of laser pulse as it passes through the input quartz window of the cell and, possibly, during self-focusing in argon due to the instantaneous Kerr effect.

Indeed, the 25-mm inlet window of the cuvette is thick enough to introduce a sufficiently large nonlinear phase into the optical pulse and cause its spectral broadening. According to our estimates, the B-integral calculated for the initial pulse in the quartz plate is on the order of 3 radians, which is a sign of an active pulse SPM due to Kerr effect. According to the data in Fig. 2d, the bandwidth of the pulse spectrum at $t = 0$ s at the cell output is $\Delta\lambda \approx 37$ nm, which is nearly twice its initial value ($\Delta\lambda = 18$ nm).

At the same time, this spectral broadening does not indicate the presence of a plasma trail in compressed argon, which is a true hallmark of laser pulse filamentation. Indeed, based on an estimate using the phenomenological Marburger formula [23], the transverse collapse (self-focusing) distance $z_K$ for a Gaussian optical beam with 200 GW power and 2.5 cm diameter in argon at 40 atm is approximately 11 m. This value, even when accounting for the non-ideality of the real beam's transverse profile ($M^2 = 1.3$), significantly exceeds the length of the optical cell used. Consequently, filamentation of the beam did not occur. Note, that in calculating $z_K$, we applied linear scaling of the medium's Kerr nonlinearity coefficient with pressure [17, 24]:

$$n_2(p) = \beta n_2(p_0), \qquad (2)$$

where $\beta = p/p_0$ is the reduced pressure, and $n_2(p_0) = 9.7\cdot10^{-20}$ cm$^2$/W [25].

After starting and during the pressure release in the cell, as shown in Fig. 2d, an increase in the spectral width of the radiation is observed, which ceases only at the moment of valve closure ($t = 11$ s). At this stage, the spectral width parameter $\Delta\lambda$ reaches values of approximately 80 nm. Furthermore, the pulse spectrum shape itself undergoes a notable transformation, developing a pronounced blue supercontinuum wing. This spectral feature clearly indicates the emergence of plasma regions inside the cell. This behavior suggests that under conditions of active gas outflow from the nozzle, strong turbulent flows arise inside the gas cell. These flows effectively reduce the self-focusing length to values $z_K < L$ and thereby trigger multiple pulse filamentation [7, 9]. The appearance of filaments is directly visible in the beam image presented in Fig. 2a, where multiple bright polychromatic spots and spatially localized regions can be clearly identified.

After the argon pressure release in the gas cell ends, the system enters a stage of thermodynamic relaxation. During this phase, the pulse spectrum gradually narrows, the filament traces disappear from the images of beam profile, and the beam diameter decreases. The temporal duration of this relaxation stage is somewhat longer than that of the preceding pressure release stage. and approximately 20 s after the gas valve closure the laser pulse fully relaxes back to its initial parameters.

It is important to note that the discussed dynamics of the femtosecond pulse spectrum is fully preserved during the reverse procedure of pressurization into the cell. Specifically, when the pressure rapidly increases from 30 to 40 atm (not shown here), the same sequence of spectral and spatial changes is observed: initial spectral broadening, formation of filamentation structures, and subsequent relaxation to the original beam characteristics. This reproducibility confirms that the observed effects are directly linked to the transient gas dynamics and turbulence inside the cell rather than to irreversible changes in the system.

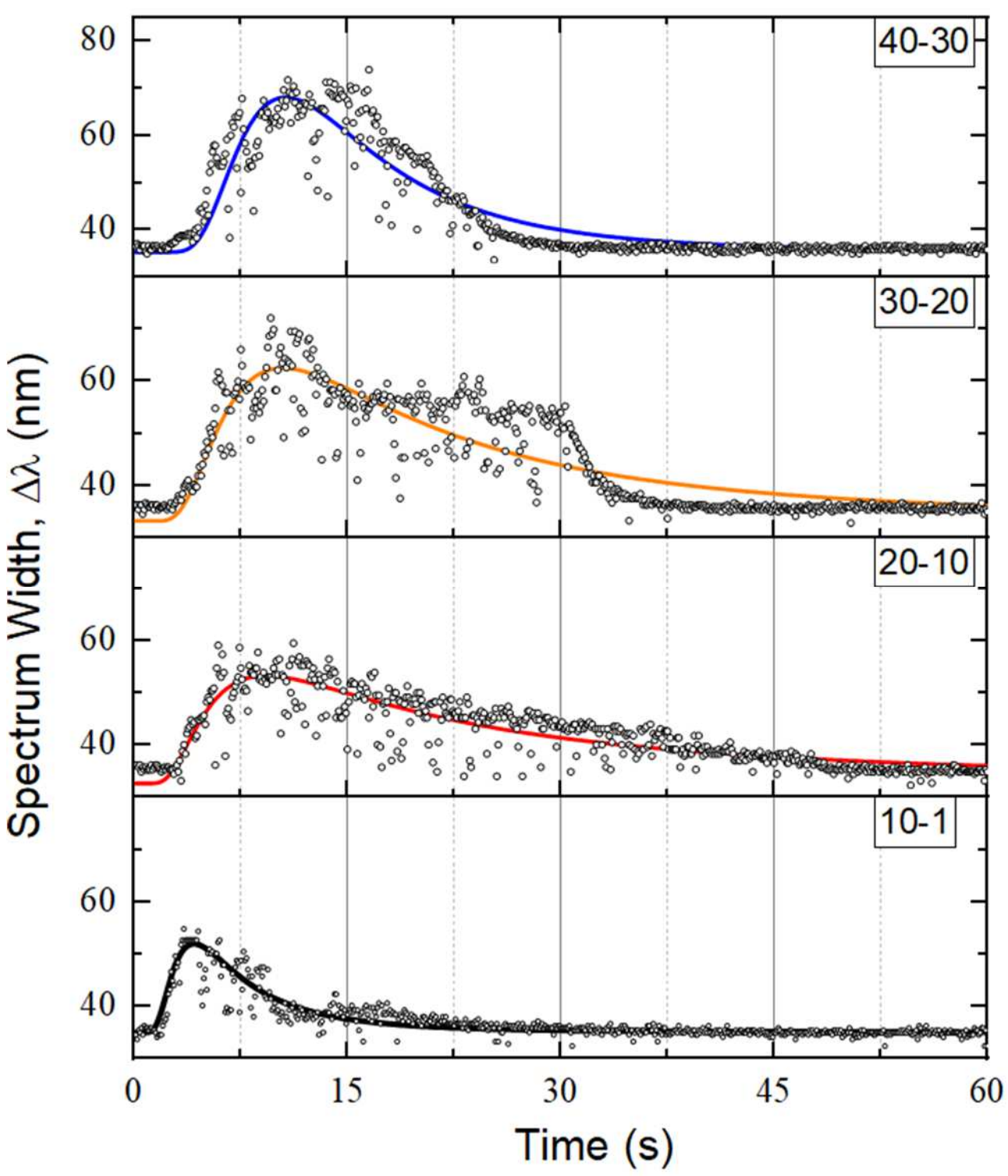

Fig. 3. Dynamics of the spectral broadening of a femtosecond laser pulse as a function of time during the controlled release and relaxation of argon pressure in the gas cell. Different pressure steps (indicated in atm, see legend) reveal distinct broadening regimes. The solid curves are least-squares fits using a Lorentzian function, illustrating the nonlinear response of the compressed gas under transient conditions.

As shown in Fig. 3, the temporal dynamics of the pulse spectral width during argon pressure release and relaxation in the gas cell reveals several key trends. The upper ($p_1$) and lower ($p_2$) pressure levels, indicated in the legend for each panel, vary between measurements, while the pressure jump, $\Delta p = p_1 - p_2$, remains constant at 10 atm. It is observed that as $p_1$ decreases, the maximum achievable spectral width $\Delta\lambda$ of the pulse also decreases. This trend suggests a reduction in the length of filaments formed inside the cell, which are seeded by internal turbulence. The turbulence, in turn, is driven by the transient gas dynamics during the rapid pressure drop. Furthermore, lowering the initial argon pressure $p_1$ in the cell generally prolongs the relaxation stage of turbulent flows. This extended relaxation time reflects the slower dissipation of turbulent energy at lower gas densities and pressures. Notably, the time required for the spectral width parameter $\Delta\lambda$ to reach its maximum value varies across the measurements, as clearly seen in the plots of Fig. 3. This variation is directly linked to the different gas outflow rates from the exhaust valve under varying initial pressure conditions $p_1$. Indeed, at higher $p_1$, the gas exits more rapidly, leading to a faster onset of turbulence and filamentation, and thus a shorter time to maximum spectral broadening. Conversely, at lower $p_1$, the outflow is slower, delaying the development of turbulence and the associated spectral changes. These observations confirm that both the spectral broadening and the filamentation dynamics are strongly governed by the interplay between pressure transients, gas flow rates, and induced turbulence inside the cell.

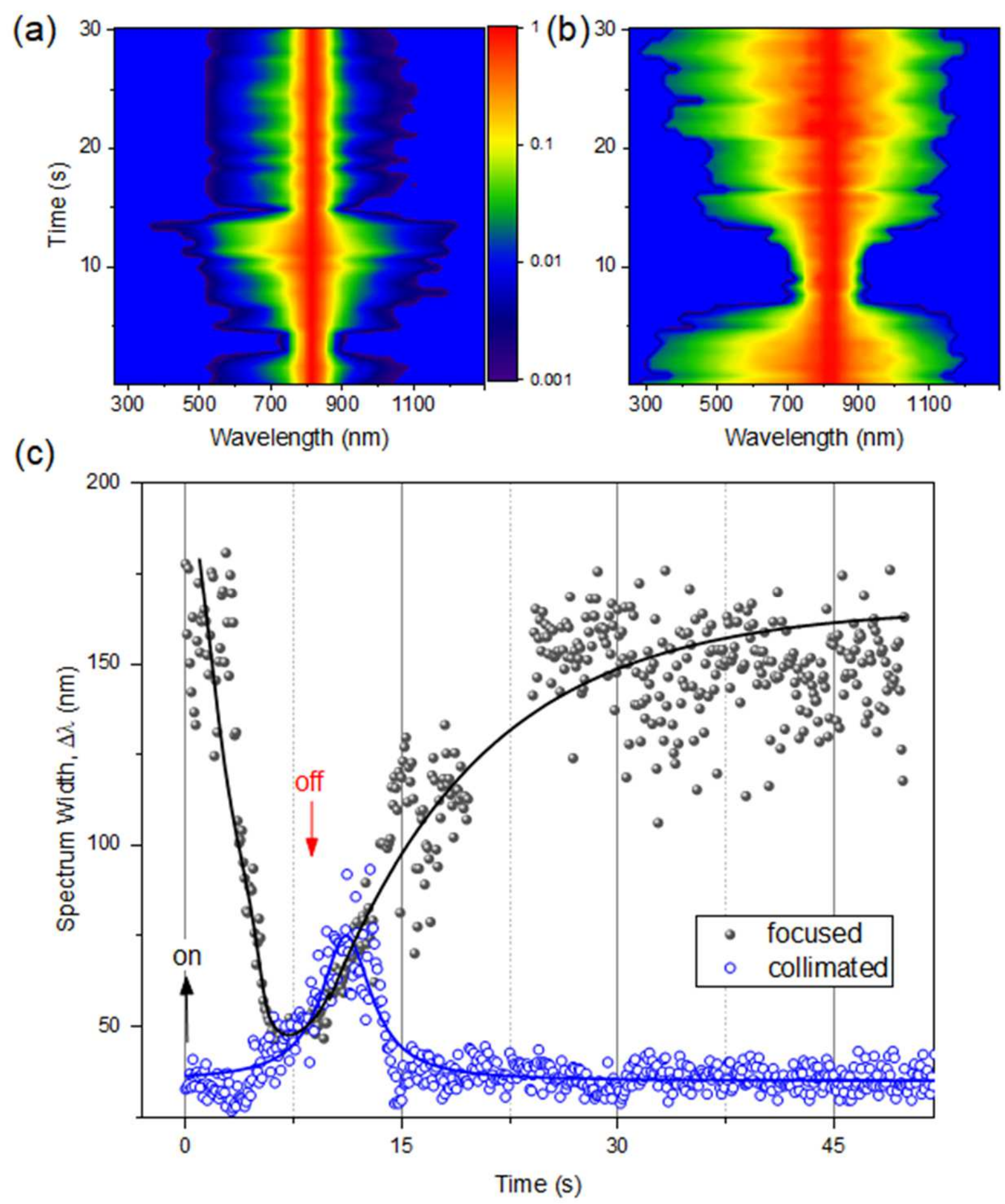

Fig. 4. Comparison of optical pulse spectra after filamentation in argon under controlled pressure release (30→20 atm) and different beam focusing geometries: (a) spectrum for a collimated 6 mm beam; (b) spectrum for a focused 25 mm beam (mirror focal length $f = 1$ m). (c) Temporal evolution of the pulse bandwidth Δλ, highlighting the influence of focusing conditions on the degree of spectral broadening during the pressure transient.

As demonstrated above, active turbulence in the gas cell induced by a rapid pressure drop accelerates the onset of optical pulse filamentation. This occurs through the generation of initial perturbations in the laser beam profile, which subsequently act as nucleation centers for filaments due to the optical nonlinearity of the medium. Consequently, in a turbulent atmosphere, multiple filamentation arises at a shorter distance from the source compared to a quiescent atmosphere. This effect manifests as an expansion of the supercontinuum spectrum at the output of the turbulent gas cell. However, turbulence not only accelerates the development of instabilities in the laser beam and triggers earlier filament formation, but also modifies the very conditions of radiation propagation. Specifically, it affects the balance between convergence effects (Kerr self-focusing) and divergence effects (plasma refraction) of the propagating beam. Under certain conditions, strong divergence of the global beam radius induced by turbulence can lead to a critical loss of intensity required to overcome diffraction during the self-focusing process. This, in turn, may cause the decay of emerging filaments. As a result, such dynamics can delay the onset or even suppress filamentation entirely [12, 26].

Formally, one can state that strong "turbulent" beam divergence increases the critical power threshold for self-focusing [9]. This implies that in weakly turbulent conditions, the initial perturbations enhance filament nucleation and reduce the filamentation onset distance. In contrast, in strongly turbulent regimes, excessive beam broadening may prevent self-focusing if the local intensity falls below the critical value needed to balance diffraction. The competition between these two opposing effects ultimately determines whether turbulence promotes or inhibits filamentation in a given experimental configuration. This dual role of turbulence, both as a seeding mechanism and as a disruptive factor, highlights the complexity of laser pulse propagation in transient gas dynamics. The observed spectral broadening (supercontinuum generation) thus reflects not only the nonlinear self-action of laser pulse, but also the interplay between turbulence-induced perturbations and the intrinsic nonlinear response of the gas medium.

A similar effect was observed in our experiments when we directed a pre-focused pulse into a gas cell filled with argon compressed to 30 atm, followed by an emergency pressure release down to 20 atm. The comparison of the temporal dynamics of the spectra and the spectral half-width parameter $\Delta\lambda$ for collimated (6 mm beam diameter, 2.8 mJ energy) and mirror-focused (25 mm beam diameter, 1 mJ energy, focal length $f = 1$ m) femtosecond radiation is presented in Figs. 4a–c. It is evident that the turbulence initiated by the pressure release inside the cell affects collimated and converging radiation in different ways. At the moment the valve opens (marked as "on" in Fig. 4c) and immediately thereafter, the collimated beam exhibits virtually no change in its spectral composition. In contrast, the pre-focused pulse, which is already in the filamentation regime at the time of valve opening (spectral bandwidth $\Delta\lambda \approx 170$ nm), shows an immediate narrowing of its spectrum. This indicates that filamentation of the focused radiation at least shifts further away from the geometric focus, towards the output window of the cell, due to the influence of turbulence. Consequently, the effective length of nonlinear spectral transformations of the pulse, which determines the width of the supercontinuum spectrum, is reduced.

By the time the valve closes (marked as "off" in Fig. 4c), the focused optical beam exhibits a spectral width of approximately 50 nm, suggesting that filamentation within the cell has effectively ceased. In contrast, the spectrum of the unfocused pulse at this moment is at its broadest, indicating active filamentation. During the subsequent stage of thermodynamic relaxation of the gas in the cell, the spectral composition of the radiation gradually recovers to its initial parameters. Notably, the focused pulse recovers significantly more slowly than the collimated beam. This difference clearly demonstrates the strong influence of turbulent gas flows on filamentation dynamics under conditions of external beam focusing. The slower recovery of the focused pulse reflects the prolonged dissipation of turbulence-induced perturbations, which continue to affect beam spatial and spectral properties even after the pressure transient has ended.

## 4. Fluid dynamics simulation of shock gas outflow

As demonstrated in the previous section, the behavior of intense femtosecond radiation during a rapid pressure drop in the surrounding gas is generally characterized by two effects: a shift of the beam centroid and a change in the conditions of its filamentation (either stimulation or damping). The most probable physical cause of these effects is the excitation of strong turbulent flows of dense gas near the outlet valve of the steel cell. Below, we will substantiate this assertion using simulation results based on the numerical solution of fluid dynamics equations.

Fluid dynamics modelling of compressed gas outflow from an optical cell into the surrounding air is carried out by addressing the classical problem of gas outflow from a cylindrical volume (a steel tube) at a certain pressure $p_0$ = 1 atm. It is known that, under adiabatic conditions, depending on the pressure ratio $\beta = p_1/p_0$, the process of unloading the gas cavity can proceed in either a critical or a subcritical regime. In the former case, gas outflow occurs at a constant critical pressure ratio $\beta_{cr} = \left[2/(\gamma+1)\right]^{\gamma/(\gamma-1)}$, where $\gamma$ is the adiabatic exponent of the gas (for monatomic gases, $\gamma$ = 1.67). In the subcritical regime, the pressure difference continuously decreases. Worthwhile noting, estimating the velocity and mass flow rate of the outflowing gas in both regimes is not particularly challenging (see, e.g., [27]). However, calculating the actual gas dynamics inside the cell is only feasible using computational methods, specifically by solving the Navier-Stokes equations.

In the present case, we assume $p_1$ = 20 atm, so the boundary of the critical regime is given by: $p_{cr} = p_1\beta_{cr} \approx 9.7$ atm > $p_0$. Consequently, the outflow begins and proceeds at a constant velocity equal to the speed of sound in argon (~ 319 m/s). For a single-phase isothermal flow of a compressible medium, the mass and momentum conservation equations take the following form [28]:

$$\frac{\partial\rho}{\partial t} + \frac{\partial\rho u_i}{\partial x_i} = 0 \tag{3}$$

$$\frac{\partial\rho u_i}{\partial t} + \frac{\partial\rho u_i u_j}{\partial x_j} = -\frac{\partial p}{\partial x_i} + \frac{\partial}{\partial x_j}\left[\mu\left(\frac{\partial u_i}{\partial x_j} + \frac{\partial u_j}{\partial x_i}\right) - \frac{2\mu}{3}\frac{\partial u_k}{\partial x_k}\delta_{ij}\right] \tag{4}$$

Here, $\rho$, $u$, $p$ are the density, velocity components and gas pressure, $x_{i,j,k}$ are spatial coordinates, $\mu$ is the dynamic viscosity of the medium, $\delta_{ij}$ is the Kronecker symbol. In Eq. (4), the viscous Reynolds stress tensor (term in square brackets) is written in the Stokes formulation for a Newtonian fluid when the relationship between stress and deformations of an elementary volume of a medium obeys a linear law. In addition, for simplification, the term that takes into account the effect of gravity is omitted here.

In simulation the problem under discussion, however, we solve the system of equations (3) – (4), but the so-called Reynolds-averaged Navier-Stokes (RANS) equations [28]. These equations are derived under the assumption that random turbulent fluctuations of thermodynamic quantities are small compared to their mean values. To close the RANS equations, we employ the standard $k$-ε model of turbulent viscosity [29], which accounts for the non-locality of turbulent strain transport in the flow. This turbulence model relies on several assumptions, the most important of which is that the flow Reynolds number, $\mathrm{Re} = \rho u L/\mu$ (where $L$ is a characteristic spatial scale), must be sufficiently large. It is also essential that turbulence be in equilibrium within the boundary layers, meaning that the rate of vortex generation must, on average, equal the rate of their dissipation.

RANS equations are solved numerically using the finite volume method (FVM), implemented in the freely available computational fluid dynamics (CFD) OpenFOAM software [30]. The geometry of the problem is shown in Fig. 5. To accelerate the computation and reduce the required computational resources, the length of the optical cell, which is modelled as a cylinder with impermeable walls, is reduced to 50 cm. At the same time, the cell diameter, as well as the location and diameter of the outlet nozzle, were kept consistent with the real experimental setup. We realize, that by the reducing the longitudinal tube dimension we change the gas dynamic parameters of our cylindrical cell. However, in our approach, we do not try to quantitatively compare the experimentally observed dynamics of gas in a cell with a computer model. Our goal is to show what types of vortex motions may occur inside a pressurized gas cavity under experimental conditions. In addition, even with accurate modeling of a full-size cell, it is difficult to take into account the actual manufacturing features of the exhaust nozzle and the side flanges of the cylindrical pipe, which can also have a significant effect on the gas motion.

Spatial discretization is performed by generating a tetrahedral mesh with a maximum edge length of 3 mm, resulting in approximately $6{\cdot}10^5$ mesh elements. Standard no-slip boundary conditions are applied at all solid boundaries except the nozzle exit plane; these conditions set the tangential component of the relative gas velocity $u$ at the cell walls to zero. At the nozzle exit, a normal flow condition was imposed, i.e., the tangential component of the stress tensor was set to zero: $\mu\,\partial u_t/\partial|\mathbf{n}| = 0$.

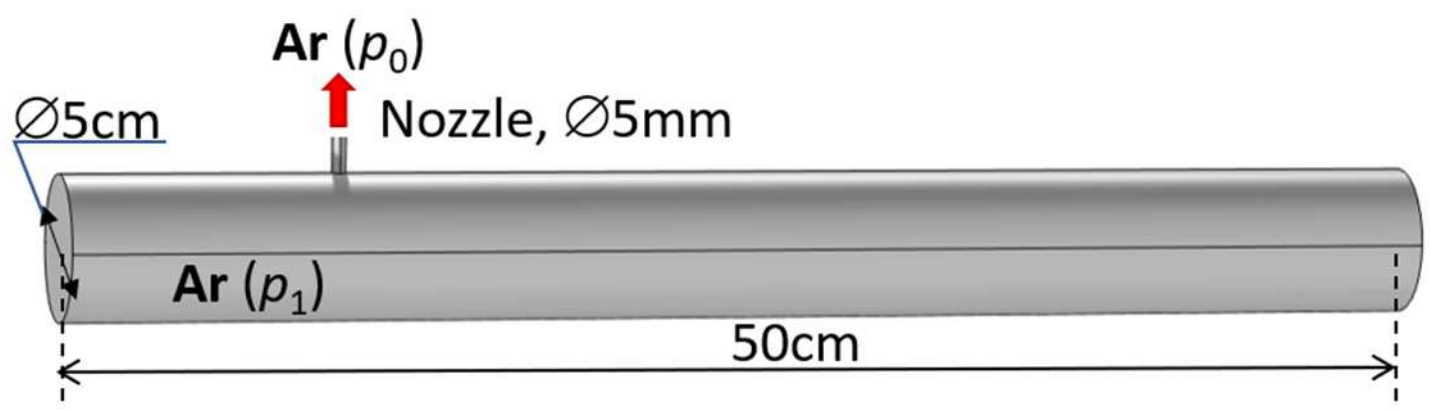

Fig. 5. Geometric diagram of the fluid dynamics setup for CFD simulation the outflow of argon (Ar) from a cylindrical high-pressure chamber ($p_1$) to the surrounding low-pressure environment ($p_0$, with $p_0 < p_1$).

The problem under consideration is divided into sequential time stages. In the first of them, the compressed gas actually flowed freely out of the cuvette until a certain average volume pressure $p_2 < p_1$ (as a rule, $p_2 \approx 10$ atm) was reached. Then, in the second stage, the nozzle is closed, and the gas flow inside the cuvette relaxed to an equilibrium state, when the average value of the gas velocity $u$ became close to zero.

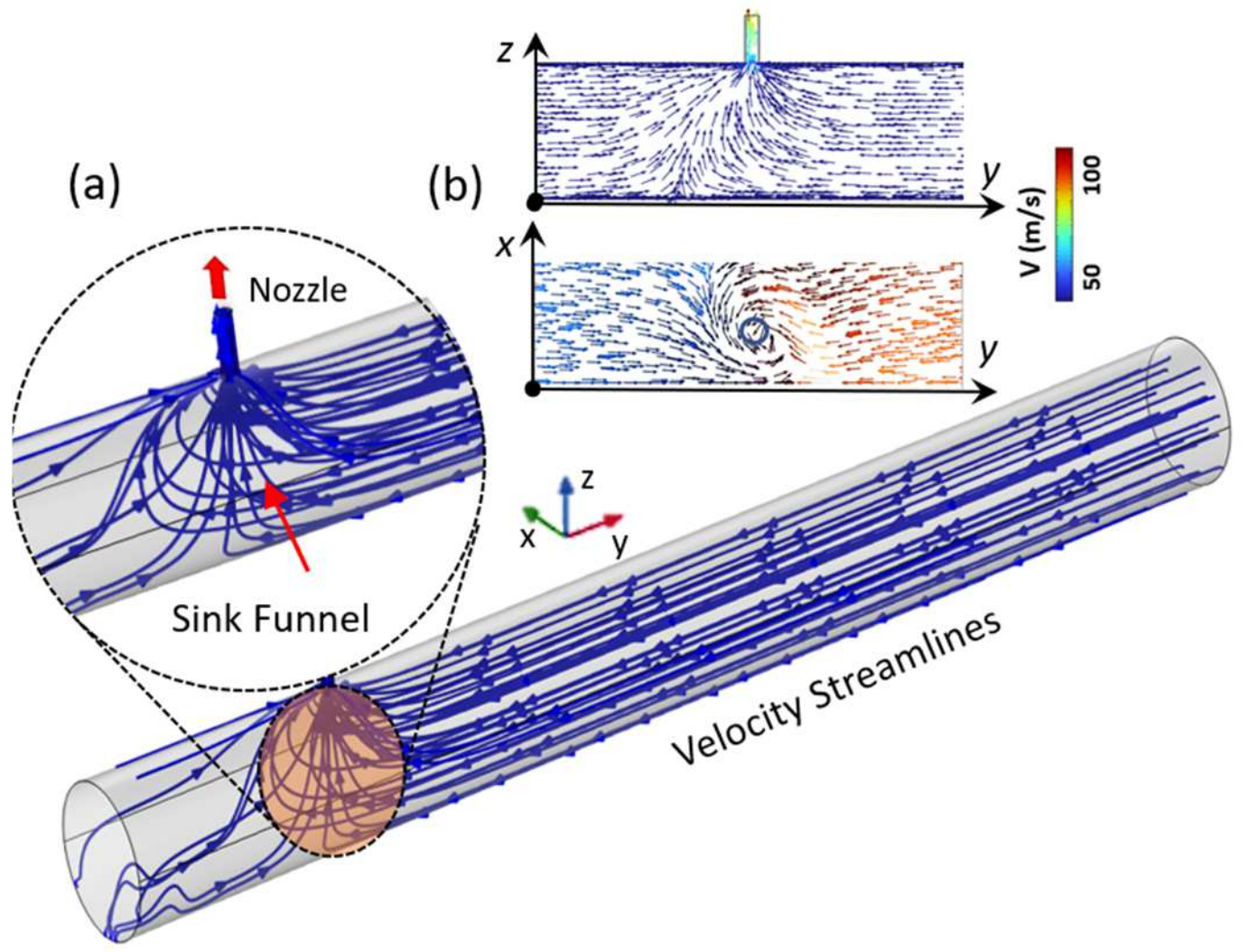

Fig. 6. CFD results for pressurized gas outflow from the optical cell volume. (a) Three-dimensional streamlines illustrating the spatial structure of the gas velocity field $u(x,y,z)$. (b) Two-dimensional vector maps of velocity magnitude and direction in two perpendicular axial cross-sections intersecting the nozzle region, providing insight into the flow anisotropy and jet formation.

The visualization of the compressed argon outflow stage from the cell volume is shown in Fig. 6. Here, the spatial distribution of the flow velocity $\mathbf{u}$ is plotted for a time instant $t = 0.8$ s from the start of the process, both as (a) three-dimensional streamlines and (b) velocity vector maps in two perpendicular cross-sections of the cell aligned along its axis in the $z$-$y$ and $x$-$y$ planes. It should be noted that we modelled a scenario with a shorter cell. Therefore, all stages of pressure release and relaxation occur over significantly shorter timescales compared to the actual experiments.

As clearly seen in Figs. 6a–b and the close-up inset, a stable vortex gas motion in the form of a sink funnel forms in the region adjacent to the nozzle. The dense gas, which moves uniformly and nearly laminarly at a speed of approximately a few meters per second from the cell ends, then spirals into this funnel, accelerates to velocities of tens of m/s, and is finally directed towards the nozzle. At the nozzle exit plane, the peak velocity can reach the speed of sound (319 m/s in argon). The simulation revealed that as long as the average pressure inside the cell exceeds the critical level $p_{cr}$, the barometric release stage of the gas cell proceeds in a fairly stable velocity regime. During this stage, a single large cyclonic vortex funnel forms, accompanied by several smaller vortex regions near the outlet nozzle.

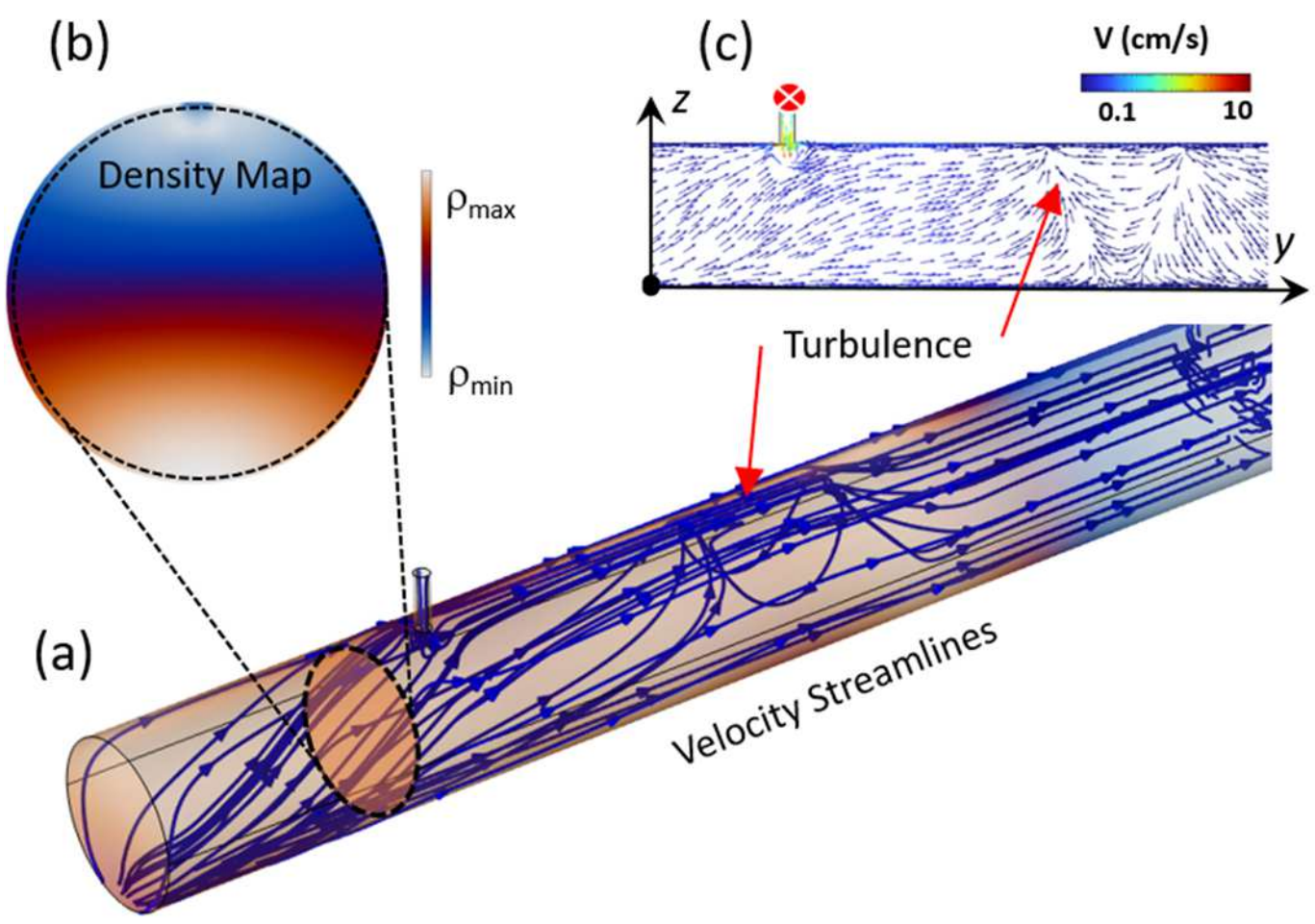

Fig. 7. Visualization of the post-outflow gas relaxation dynamics inside the cell. (a) Three-dimensional streamlines depicting the gas velocity field $u(x,y,z)$. (b) Spatial map of gas density distribution in a transverse plane. (c) Two-dimensional vector representation of the velocity field in an axial cross-section located in the vicinity of the nozzle.

At the time instant $t = 2$ s, the gas valve is closed and the cell is sealed. This marks the beginning of the argon relaxation stage, during which all thermodynamic parameters gradually relax over time towards their equilibrium values. As can be seen in Fig. 7a and *c*, at this stage the gas motion inside the cavity drastically reduces its velocity to a few or even fractions of cm/s. However, at the same time, the flow loses its stratified character and becomes highly turbulent. This circumstance leads to a non-uniform distribution of gas pressure inside the cell and, in particular, to the formation of regions with reduced density in the upper part of the tube, as shown in Fig. 7b.

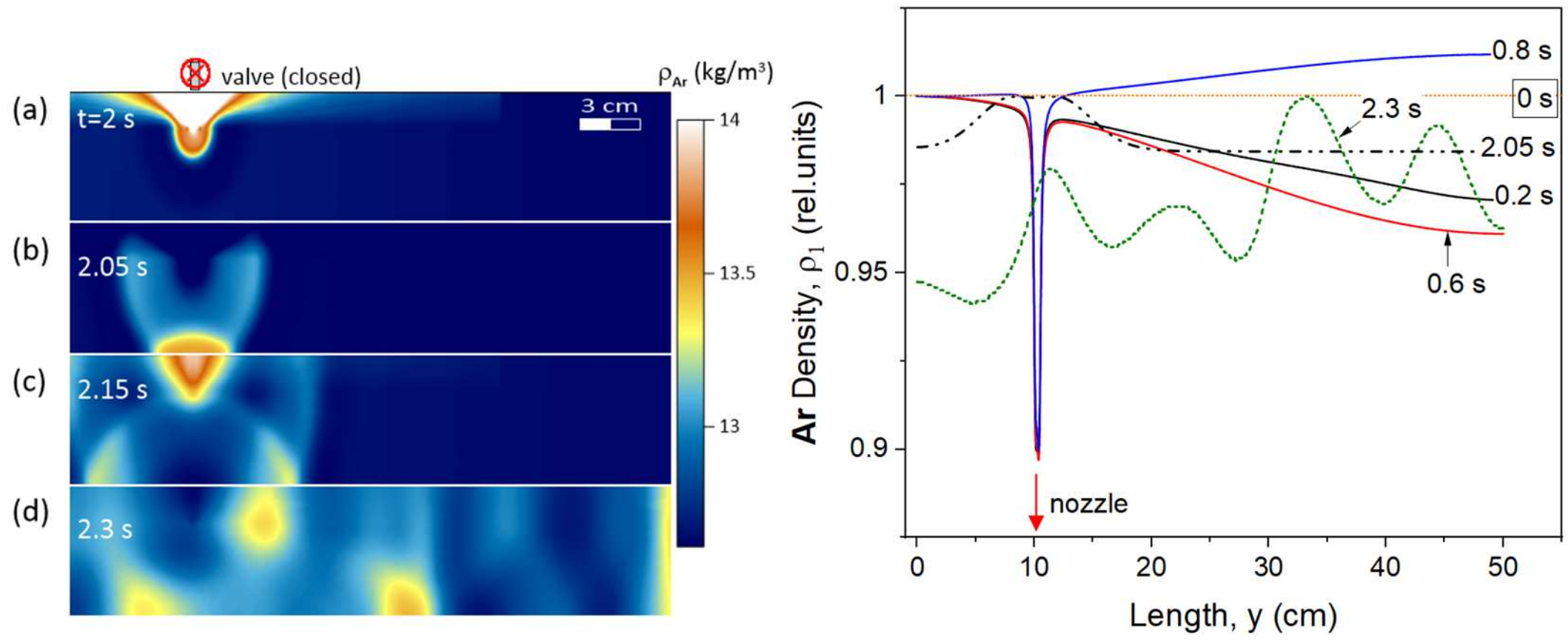

Fig. 8. Time-resolved density evolution during gas relaxation after nozzle closure. (a–d) Spatiotemporal maps of argon gas density $\rho_{Ar}$ in the nozzle vicinity, capturing the decay of pressure inhomogeneities. (e) Temporal evolution of the normalized gas density (scaled to peak) along the symmetry axis of the cell, demonstrating the approach to thermodynamic equilibrium.

The relaxation dynamics of argon are displayed in greater detail in Figs. 8a–d. It can be seen that almost immediately after the valve closure, a hemispherical region of elevated pressure forms at the top of the cell due to a gas backflow, as shown in Figs. 8a–b. This high-pressure region propagates downward and reflects off the lower surface of the tube (Figs. 8b–c). Subsequently, this pressure wave moves into the lateral undisturbed regions of the cell (Fig. 8d), giving rise to alternating zones of increased and decreased density that is a characteristic feature of turbulent flow.

The gas density normalized to its maximum value, $\rho_1 = \rho/\rho_{max}$, along the cell axis at various time instants is presented in Fig. 8e. This plot reveals that during the pressure release stage ($t$ < 2 s), a stable density dip with an amplitude of approximately 10% is observed in the nozzle region. The transient self-similarity of the density profile in this region indicates a steady flow regime, as previously noted in Fig. 6. In the remainder of the cell, periodic density waves occur due to the reflection of gas flows from the ends of the cylindrical volume. During the relaxation stage ($t$ = 2.05 and 2.3 s), the density profile ceases to be regular, and its specific shape is determined by the rate of turbulent and diffusive mixing of gas volume parts.

The argon density dependencies presented in Fig. 8 illustrate how an optical beam will behave as a whole when propagating through the cell during and immediately after the pressure release. Indeed, let us employ the Gladstone-Dale relation, which connects the refractive index of the medium $n$ with its density $\rho$ under isothermal conditions: $n - 1 = K\rho$, where $K$ is the Gladstone-Dale constant (a material-specific coefficient). After eliminating $K$, one can obtain the following relation:

$$\frac{n(\rho)-1}{n_0-1}=\frac{\rho}{\rho_0}, \tag{5}$$

where $n_0 = 1.000284$ and $\rho_0 = 1.66$ kg/m$^3$ are the argon parameters at normal conditions ($p_0$ = 1 atm, 293 K). Obviously, the low-density regions formed near the outlet nozzle of the cell will act on the optical beam as defocusing lenses, which will lead to refraction and beam tilt from the axis to the bottom of the tube, as observed in the experiment.

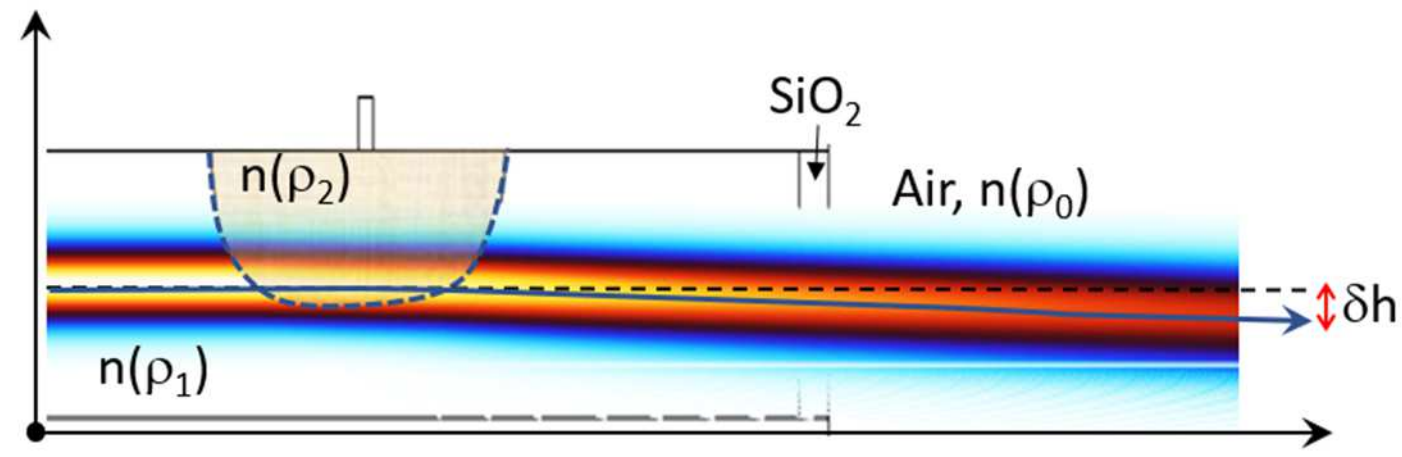

Fig. 9. Simulation of optical beam refraction caused by a gas density "bubble" using geometrical optics ray tracing.

This phenomenon can be readily illustrated within the framework of geometrical optics by ray tracing through a homogeneous medium containing a local region with a reduced refractive index. An example of such ray tracing is shown in Fig. 9 for a bundle of rays with a Gaussian intensity profile propagating inside a cylindrical tube with a half-ellipsoidal low-density cavity ("bubble") located near the exhaust valve. According to the calculation conditions, the refractive indices in the undisturbed argon and in the "bubble" corresponded to a pressure ratio $\beta = p_2/p_1 = 0.9$. The cell additionally featured an output quartz window. As seen, the vertical walk-off $\delta h$ of the optical beam center of gravity after the beam exits the cell and propagates 30 cm in air amounts to approximately 8 mm. It is evident that the vertical beam walk-off will increase with the propagation distance.

## 5. Conclusions

In conclusion, we have examined the transformation patterns of high-power femtosecond optical pulses propagating in the self-focusing and filamentation regime in gaseous argon contained in a two-meter steel cylindrical cell at elevated pressure up to 40 atm. Specifically, we investigated the effect of gas pressure shock-drop on the spectrum of the output laser radiation. Experimentally, we established that the transient optical pulse spectrum undergoes strong changes, demonstrating a multiple broadening throughout the entire time span of the pressure drop and a subsequent narrowing to the initial value (~37 nm) during the relaxation of argon pressure. The amplitude of this spectrum broadening can reach 80 nm and is proportional to the initial gas pressure. Furthermore, the spectral shape of the pulse itself changes, acquiring a pronounced blue supercontinuum wing.

During the pressure drop stage, a significant downward displacement of the optical beam centroid towards the bottom of the cell is also recorded. CFD simulations reveal that the likely physical cause of these processes is the excitation of strong turbulent gas flows in the region of the cell exhaust valve. This effect acts on the transmitted optical radiation similarly to atmospheric turbulence causing numerous flow inhomogeneities in the optical beam profile and shifting the onset of pulse filamentation closer to the beginning. However, for focused filamentation (via focusing mirror), this induced internal turbulence has the opposite effect, as it prevents the pulse from self-focusing and prevents filamentation inside the cell.

Importantly, the observed patterns persist under shock gas injection into the cell, i.e., during a pressure rise of comparable amplitude. This finding can serve as the basis for developing effective methods to control the supercontinuum spectrum via filamentation of high-power ultrashort laser pulses in high-pressure gas cells under conditions of periodic shock pressure drop-and-rise cycles. The main advantage of this technique is that there is no need to use expensive focusing optics with high radiation resistance, which makes it possible to use high-power wide-aperture laser beams to generate a supercontinuum and, accordingly, obtain powerful broadband radiation with high coherence. It is worth noting that the supercontinuum formed during the filamentation of a fs-pulse is a completely coherent radiation with a truly continuous spectrum. Its frequency band is many times higher than the bandwidth of any known lasers for atmospheric research, while maintaining the high-power characteristic of typical TW-level fs-lasers.

At the same time, in practical applications where broadband coherent light sources are required, sudden fluctuations in pulse parameters can be unfavorable. Under these conditions, achieving a stable supercontinuum output signal using periodic pressure increase cycles can be a difficult task, and the proposed method is applicable only for single trigger detection tasks.

**Funding**. Russian Science Foundation (24-12-00056); Ministry of Science and Higher Education of Russian Federation (IAO SB RAS).

**Conflict of Interest.** The authors have no conflicts to disclose.

**Author Contributions.** All authors contribute equally.

**Data availability.** Data underlying the results presented in this paper may be obtained from the authors upon reasonable request.